\documentclass[pra,twocolumn,showpacs]{revtex4}
\usepackage{amsmath,amssymb,graphicx}

\newtheorem{theorem}{Theorem}

\newtheorem{assumption}{Assumption}
\newtheorem{problem}{Problem}

\newcommand{\vmat}[2]{\left[
    \begin{array}{c}
      #1\\#2
    \end{array}\right]}
\newcommand{\dmat}[4]{\left[ 
    \begin{array}{cc}
      #1&#2\\
      #3&#4
    \end{array}\right]}

\newcommand{\hmat}[2]{\left[
    \begin{array}{cc}
      #1&#2
    \end{array}\right]}

\newcommand{\R}{{\mathbb R}}
\newcommand{\C}{{\mathbb C}}

\newcommand{\trace}{{\rm tr}}

\newcommand{\Real}{{\rm Re}}
\newcommand{\Image}{{\rm Im}}
\newcommand{\tmk}{^{\sf T}}

\newcommand{\opt}{{\rm opt}}

\begin{document}


\title{Effects of Time Delay in Feedback Control of Linear Quantum
Systems}


\author{K. Nishio, K. Kashima and J. Imura}
\affiliation{Graduate School of Information Science and Engineering,
Tokyo Institute of Technology, Tokyo 152-8552, Japan.}


\date{\today}

\begin{abstract}
We investigate feedback control of linear quantum systems subject to
feedback-loop time delays. In particular, we examine the relation between
the potentially achievable control performance and the time delays, and
provide theoretical guidelines for the future experimental setup in two
physical systems, which are typical in this research field. The
evaluation criterion for the analysis is given by the optimal control
performance formula, the derivation of which is from the classical
control theoretic results about the input-output delay systems.
\end{abstract}


\maketitle

\section{Introduction}
For reliable realization of quantum feedback control, it is
indispensable to take into consideration some real-world limitations,
such as incomplete knowledge of the physical systems and poor
performance of the control devices.  Various efforts on these issues
have been undertaken in these few years, see e.g., \cite{james2004,
james2005, james2007, yamamoto2006} for the system parameter
uncertainty. Among such limitations, time delays in the feedback loop,
which happen due to the finite computational speed of {\it classical}
controller devices, are extremely serious, since their effect may
completely lose the benefit of feedback control \cite{ramon2005,
steck2006, stockton2002}. To avoid the time delays, one can think to use
the Markovian feedback control, in which the measurement results are
directly fed back \cite{wiseman1993, wiseman2002}.  However, while these
experimental simplification has been extensively studied, theoretical
ways to evaluate the effect of the time delays have not been proposed so
far.

In this paper, we investigate the effect of the time delays on the
control performance, which is defined in terms of the cost function
optimized by feedback control. This investigation provides
theoretical guidelines for the feedback control experiment. As the
controlled object, the linear quantum systems are considered. In order
to prepare the tool for the analysis, we first consider the optimal LQG
control problem subject to the constant time delay. The optimal
controller is obtained via the existing results in the classical
control theory \cite{mirkin2003}. Further, these results allow us
to obtain the formula for the optimal value of the cost.

The obtained formula enables us to examine the relation between the
optimal control performance and the time delay both in an analytical and
a numerical ways.  Then, the intrinsic stability of the systems is
dominant for the performance degradation effect.  If the system is
stable, the degradation effect converges to some value in the large time
delay limit. Otherwise, the performance monotonically deteriorates as
the delay length becomes larger. Based on this fact, we perform the
analysis stated above for several physical systems that possess
different stability properties. In addition to the controller design, we
examine the relationship between the measurement apparatus and the best
achievable performance. Based on this, we propose a detector parameter
tuning policy for feedback control of the time-delayed systems.

This paper is organized as follows. Linear quantum control systems are
introduced in the next section. In Section III, we state the control
problem for dealing with the time delay issue, and provide its optimal
solution. In Section IV, we investigate the effect of the time delay in
quantum feedback control based on two typical examples possessing
different stability properties. Section V concludes the paper.

We use the following notation. For a matrix $A=(a_{ij})$, $A\tmk$,
$A^\dagger$ and $A^\ast$ are defined by $A\tmk=(a_{ji})$,
$A^\dagger=(a_{ji}^\ast)$ and $A^\ast=(a_{ij}^\ast)$, respectively,
where the matrix element $a_{ij}$ may be an operator and $a_{ij}^\ast$
denotes its adjoint. The symbols $\Real(A)$ and $\Image(A)$ denote the
real and imaginary parts of $A$, respectively, i.e.,
$\Real(A)=(A+A^\ast)/2$ and $\Image(A)=(A-A^\ast)/2i$. All the rules
above are applied to any rectangular matrix.

\section{Linear Quantum System}
Consider a quantum system which interacts with a vacuum electromagnetic
field through the system operator
\begin{equation}
c=Cx,
\end{equation}
where $x=[q,p]\tmk$ and $C=[c_1,c_2]\in\C^{1\times 2}$. When the system
Hamiltonian is denoted by $H$, this interaction is described by a
unitary operator $U_t$ obeying the following quantum stochastic
differential equation called the {\it Hudson-Parthasarathy equation}
\cite{hudson1984} :
\begin{equation}
\hspace{5mm}
dU_t=\left[\left(-iH-\frac{1}{2}c^\dagger
	    c\right)dt+cdB_t^\dagger-c^\dagger dB_t\right]U_t,
\end{equation}
where $U_0$ is the identity operator. The field operators $B_t^\dagger$
and $B_t$ are the creation and annihilation operator processes, which
satisfy the following quantum It\^o rule:
\begin{equation}
dB_tdB_t^\dagger=dt,\ 
dB_tdB_t=dB_t^\dagger dB_t=dB_t^\dagger dB_t^\dagger=0.
\end{equation}
Further, suppose that the system is trapped in a harmonic potential, and
that a linear potential is an input to the system. The system Hamiltonian
$H_t$ at time $t$ is given by
\begin{equation}
H_t=\frac{1}{2}x\tmk Gx-x\tmk\Sigma B u_t
\end{equation}
where $u_t\in\R$ is the control input at time $t$, the system parameters
$G\in\R^{2\times 2}$ and $B\in\R^2$ are a symmetric matrix and a column
vector, and $\Sigma$ is given by
\[
\Sigma=\dmat{0}{1}{-1}{0}.
\]
Then, by defining
$x_t=[q_t,p_t]\tmk=[U_t q U_t^\dagger, U_t p U_t^\dagger]\tmk$ and by
using the commutation relation $[q,p]=i$ and the quantum It\^o formula,
we obtain the following linear equation:
\begin{equation}
\label{system}
dx_t=Ax_tdt+Bu_tdt+i\Sigma(C\tmk dB_t^\dagger-C^\dagger dB_t),
\end{equation}
where $A:=\Sigma[G+\Image(C^\dagger C)]$. Measurement processes are
described as follows. Suppose that the field observable
$e^{-i\phi}B_t+e^{i\phi}B_t^\dagger$ is measured by the perfect homodyne
detector, where $\phi\in[0,2\pi)$ denotes the detector parameter that
the experimenter can change \cite{bouten2007}. Then, the output signal
$y_t$ is obtained by
\begin{equation}
y_t=U_t^\dagger(e^{-i\phi}B_t+e^{i\phi}B_t^\dagger)U_t.
\end{equation}
The simple calculation yields the infinitesimal increment of the
observable $y_t$ as follows:
\begin{equation}
\label{output}
dy_t=(e^{-i\phi}C+e^{i\phi}C^\ast)x_tdt+e^{-i\phi}dB_t
+e^{i\phi}dB_t^\dagger.
\end{equation}
In the following section, we refer to (\ref{system}) and
(\ref{output}) as the system dynamics and the output equation,
respectively.

\section{Optimal Feedback Control}
\subsection{Input-output delay system}

As stated in the introduction, the effect of time delays is significant
in feedback control of quantum systems. Those delays are mainly
originated from the computational time for
a controller and the transition delay of signals. Thus, they should be
modelled practically as input-output delays in the feedback loop, i.e.,
at time $t$, the signal $u_{t-h_1}$ works as a control input for the
system and the information $\{y_s\}_{s\le t-h_2}$ is available in the
controller, where we assume that $h_1$ and $h_2$ are
constants. Without loss of generality, when we consider the
optimal control problem for such a system, the total delay time can be
simply put together into one input (or output) delay. Then, the system
dynamics are modified as follows:
\begin{eqnarray}
\label{delaysystem}
dx_t=Ax_tdt+Bu_{t-h}dt+i\Sigma(C\tmk dB_t^\dagger-C^\dagger dB_t).
\end{eqnarray}
Here, the real constant $h$ denotes the total time delay in the feedback
loop, i.e., $h=h_1+h_2$. Note here that $u_t$ should be determined
by $\{y_s\}_{s\le t}$.

\subsection{Optimal control performance}
We consider the optimal control problem for the system described by
(\ref{output}) and (\ref{delaysystem}).  The following system expression
is convenient for exploiting results in the classical control
theory. Let us define a quantum noise vector
\begin{equation}
w_t:=\vmat{e^{-i\phi}B_t+e^{i\phi}B_t^\dagger}{-iB_t+iB_t^\dagger}.
\end{equation}
It is shown that the quantum noise vector satisfies the following
properties:
\begin{eqnarray}
&&\langle w_t\rangle=0,\label{noise_mean}\\
&&dw_t dw_s\tmk=\left\{\begin{array}{cc}
F_\phi dt,&\hspace{-3.5mm}\text{if}\ s=t\\
0,&\text{otherwise}
\end{array}\right.\label{noise_variance}
\end{eqnarray}
where $\langle\cdot\rangle$ denotes the expectation and $F$ is the
non-negative Hermitian matrix given by
\begin{equation*}
F_\phi:=\dmat{1}{ie^{-i\phi}}{-ie^{i\phi}}{1}.
\end{equation*}
Also, we define the matrix $S_\phi:=\frac{1}{2}(F_\phi+F_\phi\tmk)$. By
substituting the terms of the field observables $B_t$, $B_t^\dagger$
with the noise vector $w_t$, we obtain the following equations:
\begin{eqnarray}
dx_t&\hspace{-1mm}=\hspace{-1mm}&Ax_tdt+B_1dw_t+B_2u_{t-h}dt,\nonumber\\
z_t&\hspace{-1mm}=\hspace{-1mm}&C_1x_t+D_{12}u_{t-h},\label{gplant}\\
dy_t&\hspace{-1mm}=\hspace{-1mm}&C_2x_tdt+D_{21}dw_t.\nonumber
\end{eqnarray}
Here, $z_t$ is an additional output signal defined to evaluate the
system performance, and $C_1\in\R^{2\times 2}$ and $D_{12}\in\R^2$ are
matrices freely tunable in controller design. The other system matrices
are defined as follows:
\begin{eqnarray*}
B_1&\hspace{-1mm}:=\hspace{-1mm}&\Sigma\ \Image\left(C^\dagger\hmat
{{\displaystyle \frac{2\exp(-i\phi)}{1+\exp(-i2\phi)}}}
{{\displaystyle \frac{2i}{1+\exp(-i2\phi)}}}\right),\\
B_2&\hspace{-1mm}:=\hspace{-1mm}&B,\\
C_2&\hspace{-1mm}:=\hspace{-1mm}&e^{-i\phi}C+e^{i\phi}C^\ast,\\
D_{21}&\hspace{-1mm}:=\hspace{-1mm}&\hmat{1}{0}.
\end{eqnarray*}

\begin{figure}
\setlength{\unitlength}{1mm}
 \begin{center}
 \begin{picture}(70,46)
  \put(0,8){\includegraphics[width=7cm,height=3.5cm]{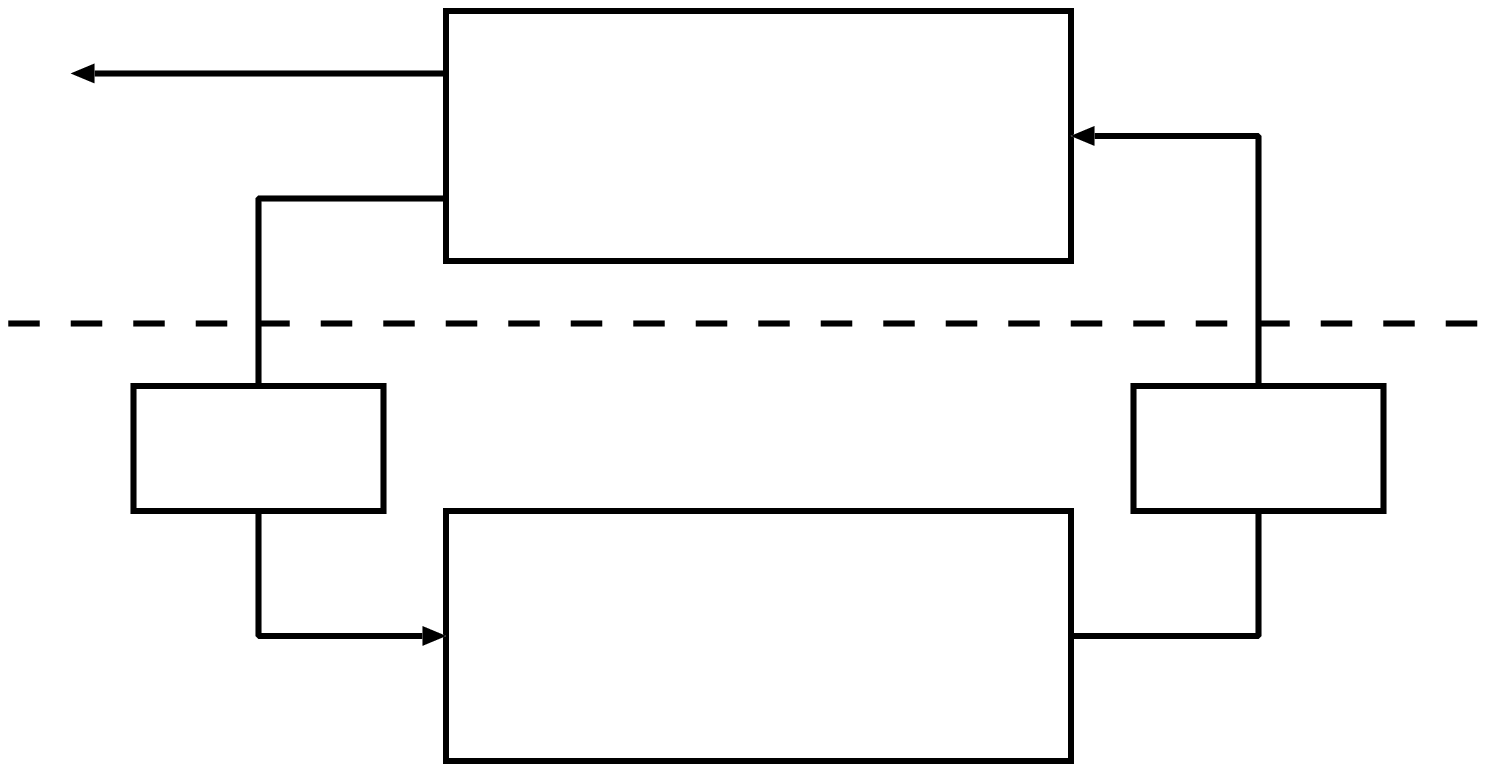}}
  \put(10,43){$z_t$}
  \put(7,30){$y_t$}
  \put(60,13){$u_t$}
  \put(35.5,39){\makebox(0,0){Plant (\ref{gplant})}}
  \put(35.5,35){\makebox(0,0){(Quantum)}}

  \put(35.2,16){\makebox(0,0){Controller}}
  \put(35.2,11){\makebox(0,0){(Classical)}}

  \put(11.7,22.3){\makebox(0,0){\scriptsize Delay \hspace{-0.5mm}$h_1$}}
  \put(58,22.3){\makebox(0,0){\scriptsize Delay \hspace{-0.5mm}$h_2$}}
 \end{picture}
 \caption{Control of quantum systems by classical controllers}
 \label{fig:loop}
 \end{center}
\end{figure}

As depicted in Fig. \ref{fig:loop}, we investigate the feedback loop
consisting of the system and a controller implemented by classical
devices, such as analogue or digital circuits.  Then, the optimal
control problem is stated as follows.
\begin{problem}
\label{problem}\rm
For the linear quantum system (\ref{gplant}), find the causal, linear
and time-invariant control law $u:\{y_s\}_{s\le t}\to u_t$ that
minimizes the cost functional
\begin{equation}
J:=\lim_{t\to\infty}\langle z_t\tmk z_t\rangle,
\end{equation}
and determine the minimum value of $J$.
\end{problem}
We make the following assumption, which is standard in the classical
control theory; see \cite{zhou1995} for the details.
\begin{assumption}\rm
\label{assumption}
\ 
  \begin{enumerate}
  \item $(A,B_2)$ is stabilizable and $(A,C_2)$ is detectable.\vspace{-2mm}
  \item For any $\zeta \in \R$,
        \[
           \quad \quad \dmat{A-j\zeta I}{B_2}{C_1}{D_{12}},
           \dmat{A-j\zeta I}{B_1}{C_2}{D_{21}}
        \]
        are row- and column-full rank, respectively.\vspace{-2mm}
  \item $E_1:=D_{12} \tmk D_{12}$ and
$E_{2,\phi}:=\ D_{21}S_\phi D_{21}\tmk$ are nonsingular.
  \end{enumerate}
\end{assumption}

As in the other results for linear quantum systems, the solution of
Problem \ref{problem} can be obtained by slightly modifying the
derivation of the classical result in \cite{mirkin2003}. Here, we only
provide the minimum value of the cost $J$, which is of importance for
the later discussion. For the specific form of the optimal controller,
see the appendix.
\begin{theorem}\rm
\label{maintheorem}
Consider Problem \ref{problem} with Assumption \ref{assumption}. Let
$X$, $Y$ be the solutions of the matrix Riccati equations $XA+A\tmk
X+C_1\tmk C_1-F\tmk E_1 F=0$ and $YA\tmk+AY+B_1S_\phi
B_1\tmk-LE_{2,\phi} L\tmk=0$ with $F:=-E_1^{-1}(B_2\tmk X+D_{12}\tmk
C_1)$ and $L:=-(YC_2\tmk+B_1S_\phi D_{21}\tmk)E_{2,\phi}^{-1}$ such that
$A+B_2F$, $A+LC_2$ are stable. Then, the optimal value of the cost
functional $J$ is given by
\begin{equation}
\label{performance}
J_{h,\phi}^\opt:=J_\phi^\opt+\int_0^h (Fe^{A\tau}L)^2 d\tau,
\end{equation}
where
$J_\phi^\opt:=\trace(B_1S_\phi B_1\tmk X)+\trace(F\tmk E_1 FY)$ is the
optimal value of $J$ when $h=0$.
\end{theorem}
Note that the existence of the Riccati solutions $X$ and $Y$ follows
from Assumption \ref{assumption}.

\section{Effect of Feedback Delay}
In the experiment of the feedback control, it is of importance to reduce
the time delay by carefully setting up the experimental devices and
achieve the best performance possible \cite{steck2006,
stockton2002}. However, some quantity of the time delay remains
in practice. In this section, we investigate how the time delay
deteriorates the optimal control performance by using the formula
(\ref{performance}). In addition to the algorithm in the controller, we
have the tunable parameter in the measurement apparatus. Thus, we do the
analysis taking the detector parameter tuning into consideration. It
should be noted that the optimal measurement technique was first
introduced by Wiseman and Doherty \cite{wiseman2005}. Their technique is
only for the delay-free systems, i.e., the optimization of the value of
$J_\phi^\opt$.

First of all, notice that the performance degradation effect is mainly
determined by the exponential term in (\ref{performance}). This means
that the degradation is largely related to the system's intrinsic
stability, i.e., the eigenvalues of the matrix $A$. Thus, it is obvious
from the exponential growth of $J_{h,\phi}^\opt$ that the unstable
system easily deteriorates as the time delay length increases, and
that their control is significantly difficult.

On the other hand, the remaining two classes, i.e., stable and
marginally stable systems, are relatively insensitive to the time delay
and it is worth to analyze them in detail.  In order to provide some
guidelines for the experiments, we analyze the two physical systems that
frequently appear in the context of quantum feedback control.
In the following, we choose the matrices $C_1$ and $D_{12}$ as
\[
C_1=\dmat{1}{0}{0}{1},\ D_{12}=\vmat{1}{1}.
\]

\begin{figure}[t]
\setlength{\unitlength}{1mm}
 \begin{center}
 \begin{picture}(70,46)
  \put(7.5,3){\includegraphics[width=5.8cm,height=3.8cm]{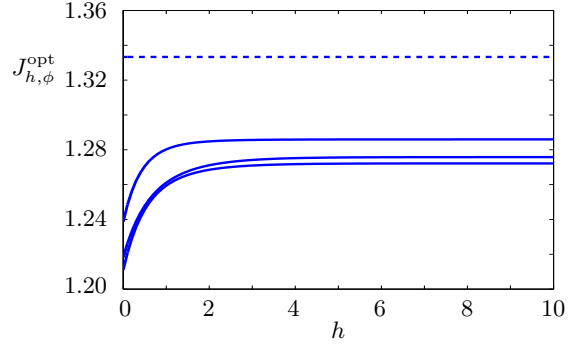}}
  \put(0,3){$1.20$}
  \put(0,12){$1.24$}
  \put(0,21.3){$1.28$}
  \put(0,30.5){$1.32$}
  \put(0,39.7){$1.36$}

  \put(7.3,0){$0$}
  \put(18.4,0){$2$}
  \put(29.7,0){$4$}
  \put(41.2,0){$6$}
  \put(52.8,0){$8$}
  \put(63.4,0){$10$}

  \put(35.5,-3){$h$}
  \put(-7,32){$J_{h,\phi}^\opt$}
 \end{picture}
 \caption{Optimal performance curves for a damped cavity with an
  on-threshold parametric down converter. A dashed line depicts the value
  of $J$ when $u_t\equiv 0$. Three solid lines correspond to the optimal
  control performances with the detector parameters $\phi=1.68$, $2.28$,
  $1.98$, from the top downwards, respectively, where $\phi=1.98$ is the
  optimal detector parameter.}
 \label{fig:damped}
 \end{center}
\end{figure}

{\bf Stable system -}
Consider a damped cavity with an on-threshold parametric down
converter. The system Hamiltonian and the coupling operator are given by
\begin{equation}
\label{parametric}
H_t=\frac{\gamma}{2}(qp+pq)-u_{t-h}q,\ \ c=\delta(q+ip),
\end{equation}
where $\gamma>0$ and $\delta>0$ are constant parameters. If they satisfy
$\gamma<\delta^2$, the system is stable. In this case, clearly,
$J_{h,\phi}^\opt$ converges as $h\to\infty$ since the real part of every
eigenvalue of $A$ is negative. When we choose the parameter as
$\gamma=1/2$ and $\delta=1$, the optimal performance curves with the
different detector parameters $\phi$ are given by Fig. \ref{fig:damped}.
It is certainly confirmed that the performance degradation converges in
the limit of $h\to\infty$ for any detector parameters.  The dashed line
depicts the value of $J$ when there is no control input field,
i.e., $u_t\equiv 0$ for any $t\ge 0$. We can see from the figure
that even in the large delay limit, the appropriate measurement strategy
significantly enhances the control performance compared to the
uncontrolled case.

\begin{figure}
\setlength{\unitlength}{1mm}
 \begin{center}
 \begin{picture}(70,46)
  \put(7.5,5){\includegraphics[width=5.8cm,height=3.8cm]{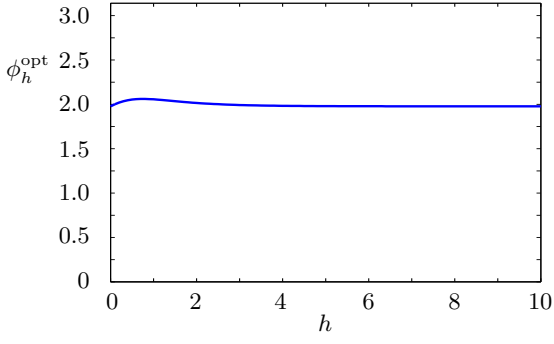}}

  \put(3.5,4.8){$0$}
  \put(1,10.7){$0.5$}
  \put(1,16.5){$1.0$}
  \put(1,22.2){$1.5$}
  \put(1,28.3){$2.0$}
  \put(1,34){$2.5$}
  \put(1,40){$3.0$}

  \put(7.3,2){$0$}
  \put(18.4,2){$2$}
  \put(29.7,2){$4$}
  \put(41.4,2){$6$}
  \put(52.8,2){$8$}
  \put(63.4,2){$10$}

  \put(35.5,-1){$h$}
  \put(-6,33){$\phi_h^\opt$}
 \end{picture}
 \caption{Optimal homodyne detector parameter v.s. each delay time
for a damped cavity with an on-threshold parametric down converter.}
 \label{fig:opt_detector}
 \end{center}
\end{figure}

In order to examine the effective detector parameter tuning, let us look
at Figs. \ref{fig:damped} and \ref{fig:opt_detector}. These show that
the optimal detector parameter hardly fluctuates over every delay
length. In fact, we can see from Fig. \ref{fig:damped} that the
difference between two performance curves around the optimal one, i.e.,
the curve corresponding to $\phi=1.98$, is very small.  Further, it is
shown that the optimal measurement strategy in the delay-free case is
perfectly the same as that in the large delay case.
\begin{theorem}\rm
Consider Problem \ref{problem} with $H_t$ and $c$ defined by
(\ref{parametric}). Let $\phi_h^\opt$ denote the detector parameter that
minimizes $J_{h,\phi}^\opt$ for the fixed delay length $h$. Then, the
following relation holds:
\begin{equation}
\lim_{h\to\infty}\phi_h^\opt=\phi_0^\opt.
\end{equation}
\end{theorem}
{\bf Proof.}~
The goal is to show that the minimal values of $J_\phi^\opt$ and
$J_{\infty,\phi}^\opt$ are achieved with the same detector parameter
value.  Firstly, we compute the value of
$\partial J_\phi^\opt/\partial \phi$. Note that
\begin{eqnarray*}
J_\phi^\opt&=&\trace(B_1S_\phi B_1\tmk X)+\trace(F\tmk E_1 F Y)\\
&=&\delta^2(x_{11}+x_{22})+\frac{1}{2}(x_{12}+1)^2y_{11}\\
&&+(x_{12}+1)(x_{22}+1)y_{12}+\frac{1}{2}(x_{22}+1)^2y_{22},
\end{eqnarray*}
where
\[
X=\left[\begin{array}{cc}x_{11}&x_{12}\\x_{12}&x_{22}\end{array}\right],\ 
Y=\left[\begin{array}{cc}y_{11}&y_{12}\\y_{12}&y_{22}\end{array}\right].
\]
Then, since $X$ is independent of $\phi$, we obtain
\begin{eqnarray}
\frac{\partial J_\phi^\opt}{\partial \phi}
&\hspace{-2mm}=\hspace{-2mm}&
\frac{1}{2}(x_{12}+1)^2\left(\frac{\partial y_{11}}{\partial \phi}\right)
\nonumber\\
&&
+(x_{12}+1)(x_{22}+1)\left(\frac{\partial y_{12}}{\partial \phi}\right)
\nonumber\\
&&+\frac{1}{2}(x_{22}+1)^2
\left(\frac{\partial y_{22}}{\partial \phi}\right).\label{Jphi}
\end{eqnarray}
Besides, the relation
\[
\frac{\partial}{\partial \phi}
(YA\tmk+AY+B_1S_\phi B_1\tmk-LE_{2,\phi}L\tmk)=0
\]
provides us with
\begin{eqnarray*}
&&\frac{\partial}{\partial \phi}\int_0^h(Fe^{A\tau}L)^2 d\tau\\
&=&\int_0^h Fe^{A\tau}\left(\frac{\partial}{\partial \phi}LL\tmk\right)
e^{A\tmk\tau}F\tmk d\tau\\
&=&\int_0^h Fe^{A\tau}\left(\frac{\partial Y}{\partial \phi}A\tmk
+A\frac{\partial Y}{\partial \phi}\right)e^{A\tmk\tau}F\tmk d\tau\\
&=&-\frac{1}{2}\left\{
\frac{(x_{12}+1)^2}{2}\frac{\partial y_{11}}{\partial \phi}
+(x_{12}+1)(x_{22}+1)\frac{\partial y_{12}}{\partial \phi}\right.\\
&&\left.+\frac{(x_{22}+1)^2}{2}\frac{\partial y_{22}}{\partial \phi}
\right\}+\frac{(x_{12}+1)^2}{4}e^{-2(\delta^2-\gamma)h}\frac{\partial
y_{11}}{\partial \phi}\\
&&+\frac{(x_{12}+1)(x_{22}+1)}{2}e^{-2\delta^2h}
\left(\frac{\partial y_{12}}{\partial \phi}\right)\\
&&+\frac{(x_{22}+1)^2}{4}e^{-2(\delta^2+r)h}
\left(\frac{\partial y_{22}}{\partial \phi}\right),
\end{eqnarray*}
where we used
\[
\frac{\partial}{\partial \phi}(B_1S_\phi B_1\tmk)=0,\ 
LE_{2,\phi}L\tmk=LL\tmk.
\]
Thus, with the attention to (\ref{Jphi}), the following equation is
obtained:
\begin{eqnarray*}
\frac{\partial J_{h,\phi}^\opt}{\partial \phi}
&=&\frac{\partial J_\phi^\opt}{\partial \phi}
+\frac{\partial}{\partial \phi}\int_0^h(Fe^{A\tau}L)^2d\tau\\
&=&\frac{1}{2}
\frac{\partial J_\phi^\opt}{\partial \phi}
+\frac{(x_{12}+1)^2}{4}e^{-2(\delta^2-r)h}
\left(\frac{\partial y_{11}}{\partial \phi}\right)\nonumber\\
&&+\frac{(x_{12}+1)(x_{22}+1)}{2}e^{-2\delta^2h}
\left(\frac{\partial y_{12}}{\partial \phi}\right)\nonumber\\
&&+\frac{(x_{22}+1)^2}{4}e^{-2(\delta^2+r)h}
\left(\frac{\partial y_{22}}{\partial \phi}\right).
\end{eqnarray*}
Hence, we obtain
\[
\frac{\partial J_{\infty,\phi}^\opt}{\partial \phi}
=\lim_{h\to\infty}\frac{\partial J_{h,\phi}^\opt}{\partial \phi}
=\frac{1}{2}\frac{\partial J_\phi^\opt}{\partial \phi},
\]
where the first equality follows from the pointwise convergence of
$J_{h,\phi}^\opt$ and the uniform convergence of
$\partial J_{h,\phi}^\opt/\partial\phi$ on the domain of $\phi$. This
completes the proof.$~\blacksquare$\\
From the discussion above, we can conclude that Wiseman's
measurement strategy (the optimal tuning for delay-free case) is valid
for the stable delay systems in that $\phi_h^\opt\approx\phi_0^\opt$ for
any $h\ge 0$.
\vspace{2mm}

{\bf Marginally stable system -}
The next system is a single particle trapped in the harmonic potential
and coupled to the probe field via the position operator. The system
Hamiltonian and the coupling operator are given by
\begin{equation}
\label{harmonic}
H_t=\frac{1}{2}m\omega^2 q^2+\frac{1}{2m}p^2-u_{t-h}q,\ \ c=q,
\end{equation}
where $m$ and $\omega$ are the mass of the particle and the angular
frequency of the harmonic potential, respectively. For this system,
the shape of the optimal performance curves is analytically calculated.
\begin{theorem}\rm
\label{main_evaluation}
Consider Problem \ref{problem} with $H_t$ and $c$ defined by
(\ref{harmonic}). Then there exist constants ${\sf A}$,
${\sf B}$ and $\theta$ such that the best achievable performance is
given by
\begin{equation}
J_{h,\opt}^\opt=J_\phi^\opt+{\sf A}h+{\sf B}\sin(\omega h+\theta).
\end{equation}
Moreover, ${\sf A}$ and ${\sf B}$ are independent of the choice of $\phi$.
\end{theorem}
{\bf Proof.}~
Notice that the Riccati solution $Y$ is dependent on the parameter
$\phi$. To make this dependence explicit, we write $Y$ as $Y_\phi$ and,
similarly, $L$ as $L_\phi$ throughout the proof. Then, by Theorem
\ref{maintheorem}, the best achievable performance is given by
\begin{equation}
\label{BAP}
J_{h,\phi}^\opt = J_\phi^\opt + \int_0^h(Fe^{A\tau}L_\phi)^2 d\tau,
\end{equation}
where $J_\phi^\opt$ is the positive constant given by
\[
J_\phi^\opt=\trace(B_1 S_\phi B_1\tmk X)+\trace(F\tmk E_1 F Y_\phi).
\]
On the other hand, direct computation yields
\begin{eqnarray*}
&&\hspace{-10mm}Fe^{A\tau}L_\phi\\
&&\hspace{-10mm}
=\sqrt{\left\{l_1^2+\left(\frac{l_2}{m\omega}\right)^2\right\}
\left\{f_1^2+(m\omega f_2)^2\right\}}\sin(\omega\tau+\theta),
\end{eqnarray*}
where $F=[f_1,f_2]$, $L_\phi=[l_1,l_2]\tmk$ and $\theta_\phi$ satisfies
\[
\tan\theta=\frac{m\omega(f_1l_1+f_2l_2)}{f_1l_2-(m\omega)^2f_2l_1}.
\]
By combining this with (\ref{BAP}), we obtain the first claim.

It should be emphasized that $\phi$ contributes to $\sf A$ and $\sf B$
only through
\begin{equation}
\label{coeff}
 l_1^2+\left(\frac{l_2}{m\omega}\right)^2.
\end{equation} 
Hence, it is sufficient to show the second claim that 
(\ref{coeff}) does not depend on $\phi$.
When defining
\[
Y_\phi=\left[\begin{array}{cc}
y_{11}&y_{12}\\y_{12}&y_{22}
\end{array}\right],
\]
simple calculation yields
\begin{eqnarray*}
&&\hspace{-5mm}(4\cos^2\phi)y_{12}^2+(2m\omega^2-4\sin2\phi)y_{12}
+4\sin^2\phi-1=0,\\
&&\hspace{-5mm}(2m\cos^2\phi)y_{11}^2-y_{12}=0.
\end{eqnarray*}
Then, we obtain the following:
\begin{eqnarray*}
&&\hspace{-4mm}l_1^2+\left(\frac{l_2}{m\omega}\right)^2\\
&&\hspace{-4mm}=(4\cos^2\phi)y_{11}^2
+\frac{4(\cos\phi\ y_{12}-\sin\phi)^2}{m^2\omega^2}\\
&&\hspace{-4mm}=\frac{1}{m^2\omega^2}
\left\{(4\cos^2\phi)y_{12}^2
+(2m\omega^2-4\sin2\phi)y_{12}+4\sin^2\phi\right\}\\
&&\hspace{-4mm}=\frac{1}{m^2\omega^2}.
\end{eqnarray*}
This completes the proof.$~\blacksquare$

\begin{figure}[t]
\setlength{\unitlength}{1mm}
 \begin{center}
 \begin{picture}(70,46)
  \put(7.5,3){\includegraphics[width=5.8cm,height=3.8cm]{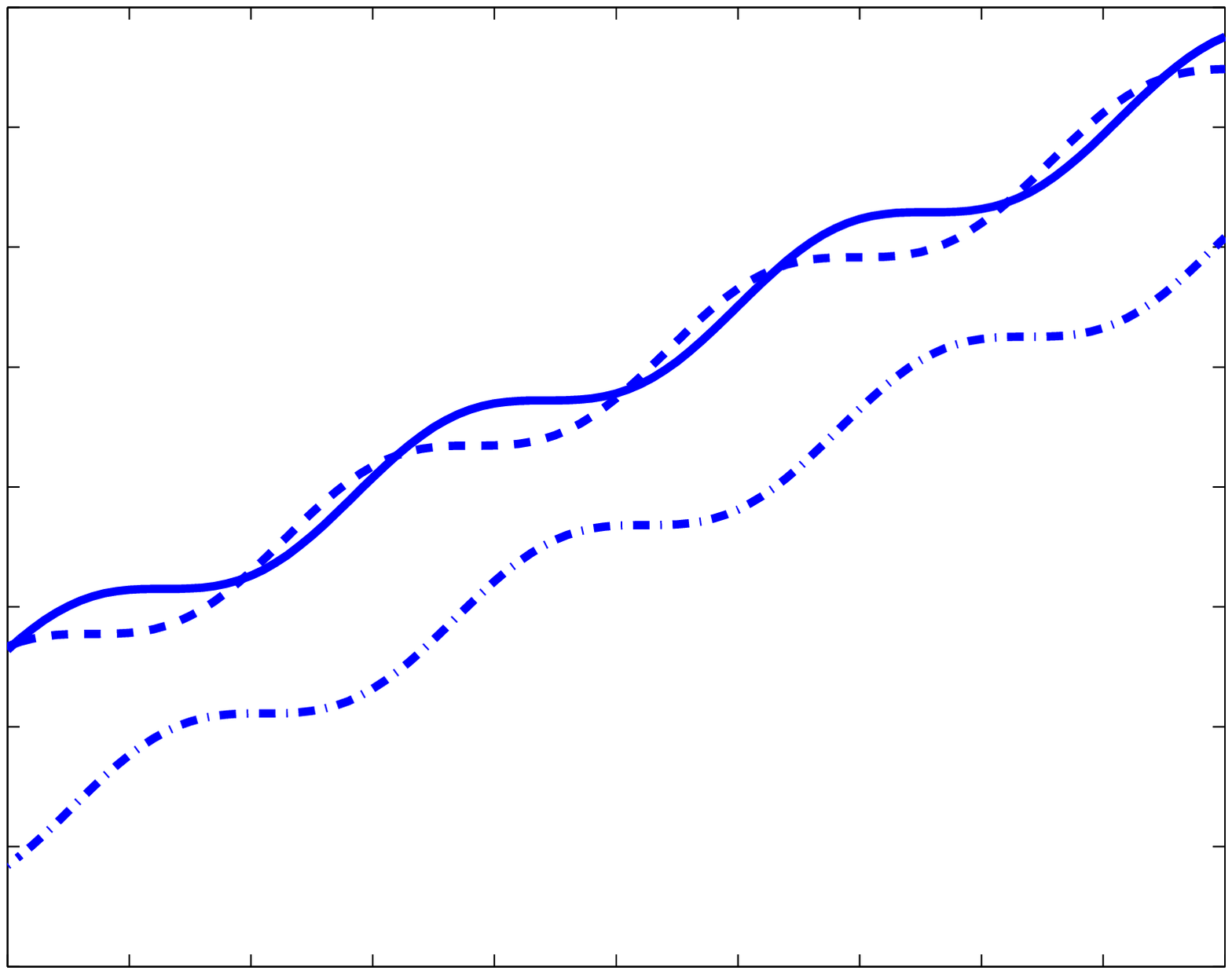}}
  \put(3,2.5){$0$}
  \put(3,7.2){$1$}
  \put(3,12){$2$}
  \put(3,16.6){$3$}
  \put(3,21.3){$4$}
  \put(3,25.9){$5$}
  \put(3,30.2){$6$}
  \put(3,34.8){$7$}
  \put(3,39.6){$8$}

  \put(7.3,0){$0$}
  \put(18.4,0){$2$}
  \put(29.7,0){$4$}
  \put(41.2,0){$6$}
  \put(52.8,0){$8$}
  \put(63.4,0){$10$}

  \put(35.5,-3){$h$}
  \put(-5,30){$J_{h,\phi}^\opt$}
 \end{picture}
 \caption{Optimal performance curves for a harmonic oscillator. A solid
  line, dashed line and chain line correspond to the optimal performances
  with $\phi=0$, $2\pi/18$, $3\pi/18$, respectively.}
 \label{fig:harmonic}
 \end{center}
\end{figure}

\noindent
Roughly speaking, the first statement says that $J_{h,\phi}^\opt$
increases linearly with the oscillation as the delay length $h$ becomes
large. This is a natural result of the fact that the matrix $A$ has only
pure imaginary eigenvalues. On the other hand, the second statement
gives us a nontrivial insight: the growth rate ${\sf A}$ and the
oscillation amplitude ${\sf B}$ are independent of the detector
parameter $\phi$.

Hence, depending on the delay length, the importance of the choice of
the measurement apparatus differs. If the delay length is small, the
improvable performance level is sensitive to the value of $h$. On the
other hand, if the system suffers from the large delay, the apparatus
adjustment is not significant, since the improvable performance level is
relatively small compared to the value of $J_{h,\phi}^\opt$. Thus, if
the time delay can be made sufficiently small, experimenters have to
adjust the detector parameter depending on the resulting delay length.
For the illustration, see Fig.\ref{fig:harmonic} for $m=\omega=1$,
which illustrates the optimal performance curve $J_{h,\phi}^\opt$.

\section{Conclusion}
In this paper, we investigated performance degradation effects due to
time delays in optimal control of linear quantum systems. The analysis
was performed by the optimal control performance formula from the
classical control theory. The obtained remarks are strongly related to
the intrinsic stability of the physical systems. In particular, we
performed intimate evaluations for two typical systems with different
types of stability. These results are expected to give useful guidelines
for the future experiments.

\appendix
\section{Optimal feedback controller for Problem \ref{problem}}
Consider Problem \ref{problem}. With the same notation as that in
Theorem \ref{maintheorem}, the optimal feedback controller is given by
\begin{eqnarray}
d\hat{x}_t&=&(A+B_2F+e^{Ah}LC_2e^{-Ah})\hat{x}_t dt\nonumber\\
&&-e^{Ah}L(dy_t+\pi_t dt)
\end{eqnarray}
\begin{eqnarray}
u_t&=&F\hat{x}_t\\
d\eta_t&=&-C_2e^{-Ah}\hat{x}_t dt+(dy_t+\pi_t dt),
\end{eqnarray}
and the finite-time integration system
\begin{equation}
\pi_t=C_2\int_{t-h}^t e^{A(t-h-\tau)}B_2 u_\tau d\tau.
\end{equation}
\vspace{0mm}
It should be noted that the implementation of this controller involves
infinite-dimensional elements. Thus, computers with the finite memory
cannot implement it in a precise sense. Fortunately, however, it is
known that the approximation method proposed in \cite{mirkin2004}
permits the control with high accuracy.



\end{document}